\documentclass[aps,prl,twocolumn,groupedaddress,amsmath,amssymb,amsfonts,graphics]{revtex4}

\usepackage{graphicx}
\usepackage{bm}
\usepackage{bbold}
\usepackage{wasysym}
\renewcommand{\vec}[1]{\mathbf{#1}}

\begin{document}

\title{Valence bond solid phases in a cubic antiferromagnet}

\author{K.\ S.\ D.\ Beach}
\affiliation{Department of Physics, Boston University, 590 Commonwealth Avenue, Boston, MA 02215}
\author{Anders W.\ Sandvik}
\affiliation{Department of Physics, Boston University, 590 Commonwealth Avenue, Boston, MA 02215}

\date{December 5, 2006}


\begin{abstract}
We report on a valence bond projector Monte Carlo simulation of the cubic
lattice quantum Heisenberg model with additional higher-order exchange
interactions in each unit cell. The model supports two different valence bond 
solid ground states. In one of these states, the dimer pattern is a three-dimensional 
analogue of the columnar pattern familiar from two dimensions. In the other,
the dimers are regularly arranged along the four main diagonals in 1/8 of the unit cells.
The phases are separated from one other and from a N\'{e}el phase by strongly first 
order boundaries. Our results strengthen the case for exotic transitions in two
dimensions, where no discontinuities have been detected at the Heisenberg 
N\'{e}el--VBS transition driven by four-spin plaquet interactions.
\end{abstract}

\maketitle

In dimension greater than one, quantum antiferromagnets with nonfrustrating, local 
interactions are generically N\'{e}el ordered at zero temperature. It is possible, 
however, for competing interactions to stabilize other singlet ground states that 
have only short range magnetic correlations. These include valence bond solids (VBS)~\cite{Read}, 
which spontaneously break the translational symmetry of the lattice, and spin 
liquids~\cite{Anderson87a}, which are featureless states having no broken symmetries.

Spin liquids came to prominence following Anderson's proposal that
a resonating valence bond (RVB) description of the doped Mott insulator might provide a route to 
superconductivity in the cuprates~\cite{Anderson87a,Lee06}.
We have since learned that spin liquids are quite unusual states of matter.
Whereas N\'{e}el and VBS phases have spin-$1$ excitations, spin liquids have
spin-$\tfrac{1}{2}$ (spinon) excitations~\cite{Kivelson87}
and topological order~\cite{Topological}.
The existence of liquid phases has been established in quantum dimer models,
but for physical spin models there is at most suggestive evidence~\cite{Fustrated}.
Because of Monte Carlo sign problems, none of the purported spin liquid models is 
amenable to exact simulation except on very small clusters.

Given the limited progress that has been made, it is worthwhile to build a repertoire of 
sign-problem-free models that exhibit quantum disordered phases.
In two dimensions (2D), a good candidate is the square-lattice Heisenberg model with a 
ring exchange term that permutes the spins around an elementary plaquet.
It is known that ring exchange can destroy magnetic order and drive the system into a 
VBS phase, both for XY~\cite{XYRing}
and fully SU(2)-invariant~\cite{Lauchli05} models.
It has been argued that the transition between antiferromagnetism and VBS 
order occurs at a special quantum critical point with deconfined spinon
excitations~\cite{Senthil04}.
Such a transition may also take place between two VBS phases
with different dimer orderings~\cite{Vishwanath04,Fradkin04}.
If this scenario holds, then it may be possible---having first indentified a deconfined
critical point---to move along some new axis in phase space
(corresponding to some additional interaction) and follow
the line of deconfined quantum critical points in the hope that
it eventually opens into an extended spin liquid phase. As it turns out, 
the transitions in the XY case are either weakly first order~\cite{Kuklov04,Spanu06} 
or possibly continuous but not of the type suggested by Senthil {\it et al.} \cite{XYtransition}. 
In the SU(2)-symmetric case, however, recent work provides compelling evidence of deconfined 
quantum criticality \cite{Sandvik06b}.

The situation in three dimensions (3D) is less well understood.
Bernier and coworkers have studied the quantum Heisenberg model on the cubic lattice 
with nearest- and next-nearest-neighbour interactions~\cite{Bernier05}.
Using an Sp($N$) generalization of the spin algebra, they 
show that there are three stable spins liquid phases for large $N$. 
These phases have short range magnetic correlations at wavevectors
$(0,0,\pi)$, $(0,\pi,\pi)$, and $(\pi,\pi,\pi)$.  It is unlikely that these phases survive in the physical
$N=1$ limit. In particular, it appears certain that the $(0,0,\pi)$
and $(\pi,\pi,\pi)$ states undergo a confinement transition to a VBS state.
Because of the nature of the precursor magnetic correlations,
the translational symmetry breaking occurs either in one lattice direction,
leading to a 3D generalization of the columnar phase
familiar from 2D, or it occurs in all three directions, leading to a ``box'' VBS.
Montrunich and Senthil have studied the Mott insulating phases of
strongly correlated bosons in 3D and have reached similar conclusions
about the possible valence bond ordering~\cite{Motrunich05}.

In this Letter we discuss two different 3D generalizations of ring exchange.
In both cases, the interaction is constructed as a 
product of four spin-spin operators in each unit cell.
In addition to the N\'{e}el phase, we find two VBS phases with 
ordering vectors $(0,0,\pi)$ and $(\pi,\pi,\pi)$. Unlike the 2D case, the 
transitions between these phases are all strongly first order.
We find no evidence of spin liquid behaviour.

\begin{figure}
\includegraphics[scale=0.96]{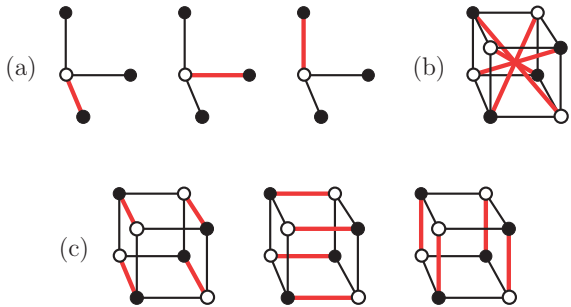}
\caption{ \label{FIG:interactions} 
The Hamiltonian contains interactions that act (a) between nearest neighbours,
(b) along the diagonals of each cube, and (c) along the edges of each cube face. 
All interactions are built from singlet projection operators between spins in opposite 
sublattices, indicated by open and filled circles.
}
\end{figure}

We consider a system of $S=\tfrac{1}{2}$ spins arranged
in a cubic lattice of even linear size $L$, subject to periodic boundary conditions.
The Hamiltonian
\begin{equation} \label{EQ:Hamiltonian}
-\hat{H} = J\sum_{-} \hat{P} + U\sum_{\boxtimes} \hat{P}\hat{P}\hat{P}\hat{P} +
V\sum_{\square}  \hat{P}\hat{P}\hat{P}\hat{P}
\end{equation}
is expressed in terms of spin singlet projection operators
$\hat{P}_{ij} = \frac{1}{4} - \vec{S}_i \cdot \vec{S}_j$ whose
site indices are understood to range over
nearest neighbours~($-$) and 
along the main diagonals~($\boxtimes$) and
face edges~($\square$) of each cubic unit cell.
The set of allowed interactions is illustrated in 
Fig.~\ref{FIG:interactions}.

The connection to physical spin operators is made by expanding
the products $(\tfrac{1}{4}-\vec{S}_i \cdot \vec{S}_j)(\tfrac{1}{4}-\vec{S}_k \cdot \vec{S}_l)\cdots$
that appear in Eq.~\eqref{EQ:Hamiltonian}. 
The first term is a two-spin interaction,
whereas the second and third terms
are alternating-sign linear combinations of two-, four-, six- and eight-spin interactions.
We also considered an interaction
$-\hat{H} \sim K \sum \hat{P}\hat{P}$ acting around the cube faces
(i.e., square plaquets symmetrized in the three orthogonal directions), but
as in the XY case~\cite{Melko05}, it is not sufficient to 
disrupt the N\'{e}el order---at least not in the
$K > 0$ parameter range that can be simulated.

The simulations were carried out using the valence bond projector Monte Carlo
algorithm~\cite{Sandvik05}.
In this scheme, the ground state is obtained by repeated application of the Hamiltonian to
a singlet trial state: $\lvert \psi \rangle = \lim_{n\to \infty}(-\hat{H})^n \lvert \psi^{\text{trial}} \rangle$. 
We selected as trial state a factorizable RVB wavefunction
with bond amplitudes of the form $h(r) = r^{-p}$~\cite{Liang88}. 
The exponent $p$ was determined self-consistently by 
measuring the powerlaw tail of the bond distribution in the final projected state.

\begin{figure}
\includegraphics{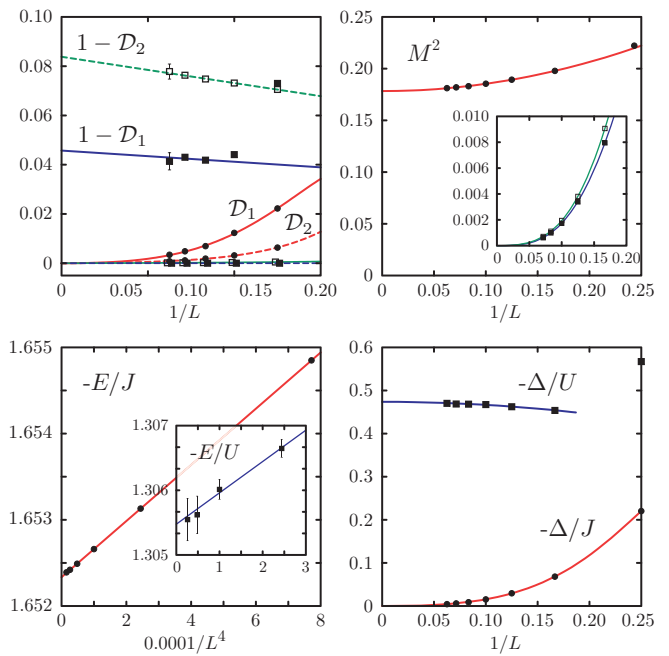}
\caption{ \label{FIG:extremes}
Point styles identify data computed in 
the extreme $J$~($\CIRCLE$), $U$~($\blacksquare$), 
and $V$~($\square$) limits.
(Top left) The VBS phases have dimer correlation functions
$\mathcal{D}_{1} = \tfrac{1}{3}\langle \hat{\mathbf{D}}^{\boxtimes}
\cdot  \hat{\mathbf{D}}^{\boxtimes} \rangle$ (solid lines)
and
$\mathcal{D}_{2} = \langle \hat{\mathbf{D}}^{\square} 
\cdot  \hat{\mathbf{D}}^{\square} \rangle$ (dotted lines)
very near saturation.
(Top right)
In the extreme $J$ limit, the sublattice magnetization 
$M = 0.4222(6)$ is within 1\% of the
spin wave theory result $M_{\text{sw}} = 0.42165$.\cite{Anderson52}.
(Bottom left)
The energy $E/J = -1.652334(5)$ is also close to
the predicted value $E_{\text{sw}}/J =-1.6455$.
In the large $U$ limit, the energy
$E/U = -0.13053(6)$ is only about 4\% below $-1/8$, which is the expectation
value of the Hamiltonian in the static valence bond pattern.
(Bottom right)
The N\'{e}el state has gapless excitations.
The extreme $U$ state has a large gap $\Delta/U = 0.474(1)$ to triplet excitations. 
The extreme $V$ state is also fully spin gapped, but the value of
the gap cannot be reliably measured using
operator-string resampling~\cite{Beach06}.
}
\end{figure}

Our main result is that the system has three phases, characterized by a nonzero sublattice
magnetization and by two forms of long range dimer order.
In the thermodynamic limit, the operators
\begin{align}
\hat{M}_a &= \sum_{\vec{r}} (-1)^{\bm{\delta}\cdot \vec{r}}S^a_{\vec{r}}\\
\hat{D}^{\boxtimes}_a &= \frac{16}{3N}\sum_{\vec{r}} (-1)^{\vec{e}_a \cdot \vec{r}}\, \vec{S}_{\vec{r}} \cdot
\vec{S}_{\vec{r} + \bm{\delta}}\\
\hat{D}^{\square}_a &= \frac{4}{3N}\sum_{\vec{r}} (-1)^{\vec{e}_a \cdot \vec{r}}\, \vec{S}_{\vec{r}} \cdot
\vec{S}_{\vec{r} + \vec{e}_a}
\end{align} 
acquire a nonzero expectation value in each of the respective phases.
Here, $a = 1,2,3$ is a label denoting the vector components and 
$\bm{\delta} = \vec{e}_1 + \vec{e}_2 + \vec{e}_3$
is a vector spanning the main diagonal of each cubic cell.
In practice, these quantities are measured on finite lattices of 
increasing size and extrapolated to the
thermodynamic limit according to
$M^2 = \lim_{L\to \infty} \langle \hat{\vec{M}} \cdot \hat{\vec{M}} \rangle$,
etc. See Fig.~\ref{FIG:extremes}.
The necessary loop estimators for the two- and four-spin operators are 
derived in Ref.~\onlinecite{Beach06}.

\begin{figure}
\includegraphics[scale=0.9]{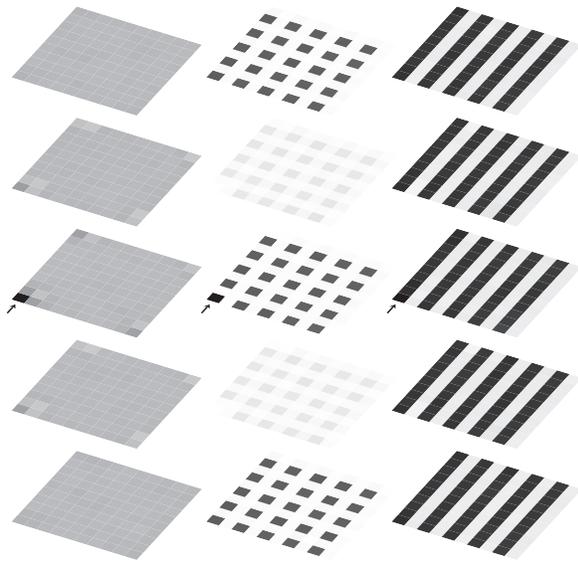}
\caption{ \label{FIG:dimers} 
A greyscale map of the average number of bonds
completely contained in each cubic unit cell; the values range from 0 (white) to 4 (black).
The measurements are made with reference to a cell
at the origin (indicated by an arrow) whose eight sites are connected by exactly 
four valence bonds.
In this example, a portion of the $L=10$ lattice is shown as a set of slices stacked vertically.
The columns correspond to the three extreme cases $J\!=\!1$, $U\!=\!V\!=\!0$ (left),
$U\!=\!1$, $J\!=\!V\!=\!0$ (centre), and $V\!=\!1$, $J\!=\!U\!=\!0$ (right).
}
\end{figure}
\begin{figure}[t]
\includegraphics[scale=0.96]{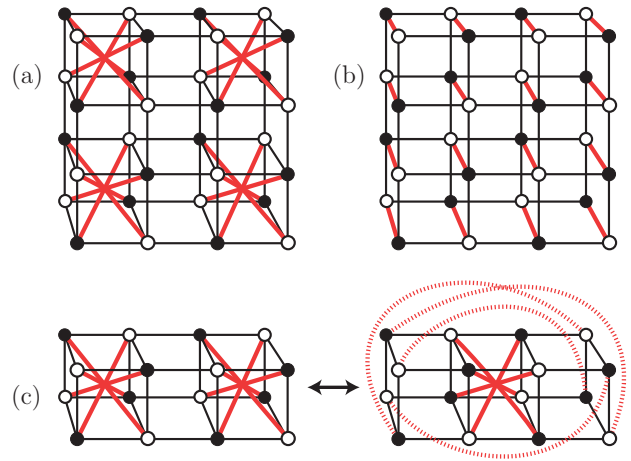}
\caption{ \label{FIG:pattern} 
(a) The $U$ interaction favours a superlattice dimer ordering in which
every eighth cell contains four valence bonds along the main diagonal.
The translational symmetry is broken in all three lattice directions.
(b) The $V$ interaction leads to a dimer configuration that
breaks the translational symmetry in one direction only.
(c) Deep in the VBS phases, fluctuations in the bond order are primarily due
to interstitial defects.
}
\end{figure}

In the extreme $J$ limit (i.e., $U,V \to 0$), the Hamiltonian reduces to the
nearest neighbour quantum Heisenberg model.
The ground state is N\'{e}el ordered ($M \neq 0$) and has a continuous
degeneracy corresponding to the spatial direction of the sublattice magnetization.
The two VBS phases are connected to the large $U$ and large $V$ limits.
In the one case, the translational symmetry is completely broken
and there is an eightfold degenerate ground state of the form
$\vec{D}^{\boxtimes} \sim (\pm 1,\pm 1,\pm 1)$. In the other, 
there is a sixfold degenerate ground state of the form
$\vec{D}^{\square} \sim (\pm 1,0,0), (0,\pm 1,0), (0,0,\pm 1)$,
which breaks translational symmetry in only one of the three lattice directions.
Our simulations indicate that the valence bond order is very strong:
the extrapolated values are
$D^{\boxtimes} = 0.98(1) \times (1,1,1)$ in the extreme $U$ limit and
$D^{\square} = 0.957(2) \times (1,0,0)$ in the extreme $V$ limit.
After equilibration,
the configurations are essentially locked into a particular dimer pattern.
The tunnelling time between energetically equivalent configurations
grows exponentially in the system size and exceeds the lifetime
of our longest simulations even for $L=6$. 

Figure~\ref{FIG:dimers} shows the dimer patterns as they occur in the 
wavefunction itself. These are imaged as follows.
For each bond configuration, we identify a unit cell
whose eight sites are connected by exactly eight valence bonds.
This cell is translated to the origin and a map is made of the
remaining cells by counting the number of valence bonds
that have both endpoints contained in the same cell.
In the N\'{e}el state, the dimer correlations do not persist beyond one or two
lattice spacings. The large~$U$ phase 
forms a spacing-2 superlattice 
in which 1/8 of the cells have valence bonds along all four main diagonals 
[Fig.~\ref{FIG:pattern}(a)]; the large~$V$ phase consists of
planar slabs of colinear bonds so that half of the cells have four bonds 
aligned in the preferred direction [Fig.~\ref{FIG:pattern}(b)].
There are weak fluctuations about these static configurations.
In the large~$U$ case, these are predominantly relocations
of a bond cluster to one of the nearest neighbour 
interstitial cells [Fig.~\ref{FIG:pattern}(c)]. For $L \lesssim 12$, 
the fluctuations interact in such a way that they preferentially align in one direction
(as evidence by the faint pattern seen in the second and fourth slices of the 
centre row of Fig.~\ref{FIG:dimers}).
As the lattice size increases, this effect diminishes and 
the rotational symmetry of the cubic lattice is restored.

\begin{figure}[!t]
\includegraphics{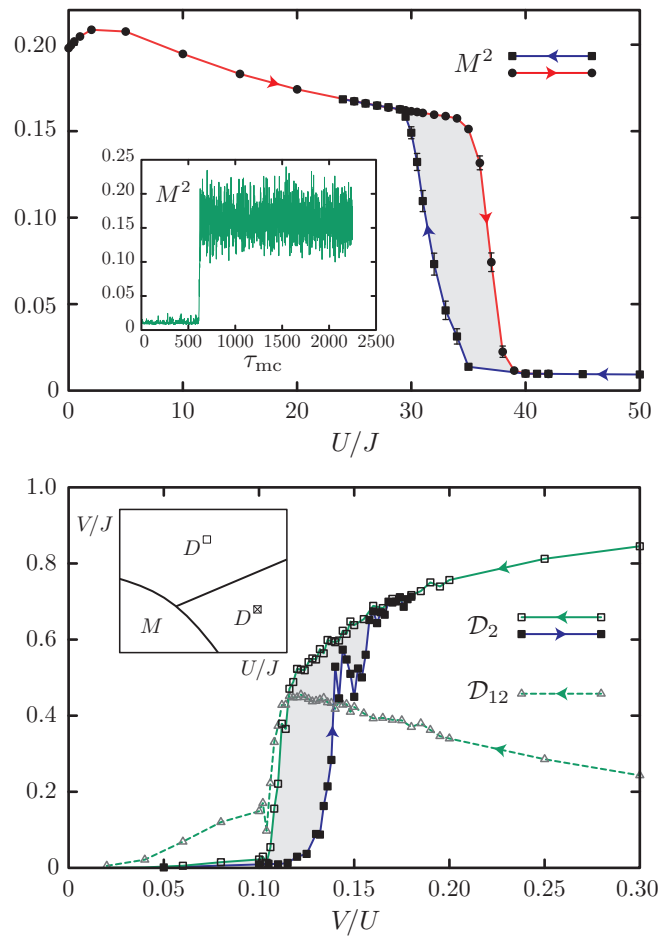}
\caption{ \label{FIG:transition} 
(Top panel)
The staggered magnetization is plotted along a $V\!=\!0$ slice of the phase diagram
for an $L\!=\!6$ system. Two data sets are shown, one generated consecutively
with $U/J$ increasing from 0 ($\CIRCLE$) and one 
with $U/J$ 
decreasing from $\infty$ ($\blacksquare$).
The inset shows the system jumping suddenly from a magnetically disordered state to
an ordered one after several hundred Monte Carlo time steps.
(Bottom panel) The dimer correlation functions $\mathcal{D}_2$
and $\mathcal{D}_{12} = \langle \hat{\mathbf{D}}^{\boxtimes}
\cdot  \hat{\mathbf{D}}^{\square} \rangle$
are plotted as a function of $V/U$ for fixed $J\!=\!0$.
Here, the system size is $L\!=\!8$.
The cube edge correlator $\mathcal{D}_2$ is computed from
both the $V/U\!=\!0$ ($\blacksquare$)
and $V/U\!=\!\infty$ ($\square$) limits. 
The mixed correlation function $\mathcal{D}_{12}$ ($\triangle$)
is shown evolving from right to left. Its discontinuity is a reliable marker
of the change in dimer order. The inset shows schematically the three phases 
separated by lines of first order transitions.
}
\end{figure}

The $M$, $D^{\boxtimes}$, and $D^{\square}$ phases are separated
by first order phase boundaries. The upper panel of Fig.~\ref{FIG:transition} 
shows the magnetization during the transition from N\'{e}el to VBS.
There are two paths shown. In one, the simulation is first performed for
$J=1$, $U=0$ and a typical configuration from this run is used as the 
starting configuration for a run at incrementally larger $U$.
A second path begins at $U=1$, $J=0$ with the results fed into new runs
at incrementally larger $J$. There is strong hysteretic behaviour beginning
at $L=6$ that becomes worse as the system size increases. 
Near the transition, the Monte Carlo sampling is
increasingly dominated by rare tunnelling events between configurations
with dimer-ordered, short range valence bonds
and those with resonating, long range bonds.
The magnetization decreases from 0.42 to $\sim 0.39$ before collapsing
abruptly.

The bottom panel of Fig.~\ref{FIG:transition} traces the dimer correlations
through the transition from VBS to VBS. Here, strong hysteresis sets in at $L = 8$. 
The simulation has difficulty moving directly between the two incompatible
 dimer orders so it accomplishes the bond reconfiguration 
 in a two-step process via some intermediate, metastable bond configuration.
 The particular pathway differs depending on the direction across
 the transition, but the general behaviour appears to be that translational 
 symmetry breaking is preceded by a process of cell substitution. This can
 be seen by looking at the mixed correlation function $\langle \hat{\mathbf{D}}^{\boxtimes}
\cdot  \hat{\mathbf{D}}^{\square} \rangle$. 
As $U/V$ decreases from $\infty$, edge cubes are replaced
by diagonal cubes while the translational symmetry remains broken at $(0,0,\pi)$
and the diagonal cubes remain uncorrelated between the occupied slabs.
The symmetry breaking transition occurs when the diagonal cubes finally organize
themselves within and between the slabs.
On the other hand, as $U/V$ increases from 0, the occupied diagonal cubes
are replaced by edge cubes of arbitrary direction, forming a 3D version
of the plaquet phase. The symmetry restoration from $(\pi,\pi,\pi)$ to $(0,0,\pi)$
comes when the edge cubes lock in a common direction.

In 2D, recent studies of a N\'eel--VBS transition in a Heisenberg model including four-spin 
interactions have shown evidence of deconfined quantum criticality. There are no signs of
discontinuities, the extracted critical exponents have reasonable values, and an emergent 
U(1) symmetry is seen explicitly \cite{Sandvik06b}. Our results presented here show
that the 3D N\'eel-VBS transition is very different, with strongly first order boundaries 
between the phases. We do not observe any spin liquid phase.

{\it Acknowledgments}---This work was supported by the NSF under grant
No.~DMR0513930.

\end{document}